\def\Bo{B_{\rm o}}
\def\bperp2{{b_\perp}^2}
\def\bpbx2{\frac{{b_\perp}^2}{{b_x}^2}}
\def\bx2{{b_x}^2}
\def\cs{c_{\rm s}}
\def\dbb{\delta B/B}
\def\'#1{\ifx#1i{\accent"13\i}\else{\accent"13#1}\fi}
\def\Ma{M_{\rm a}}

\def\Mrms{M_{\rm rms}}
\def\Ms{M_{\rm s}}

\def\va{v_{\rm A}}
\def\VS{V\'azquez-Semadeni}

\documentclass{ws-p8-50x6-00}

\begin{document}

\title{The Density Probability Distribution Function in Turbulent,
Isothermal,
Magnetized Flows in a slab geometry}

\author{Enrique Vazquez-Semadeni}

\address{Instituto de Astronom\'ia, UNAM, Apdo.\ Postal 70-264,
M\'exico, D.F.\ 04510, MEXICO\\
E-mail: enro@astroscu.unam.mx}  

\author{Thierry Passot}

\address{Observatoire de la C\^ote d'Azur, BP 4229, Nice C\'edex 4,
FRANCE\\
E-mail: passot@obs-nice.fr}  

\maketitle

\abstracts{We investigate the behavior of the magnetic pressure, $b^2$,
in fully turbulent MHD flows in ``1+2/3'' dimensions by means of its
effect on the probability density function (PDF) of the density
field. We start by reviewing our previous results for general
polytropic flows, according to which the value of the polytropic
exponent determines the functional shape of the PDF.
A lognormal density PDF appears in the isothermal
($\gamma=1$) case, but a power-law tail at either large or small
densities appears for large Mach numbers when $\gamma >1$ and
$\gamma < 1$, respectively. 
In the isothermal magnetic case, the relevant parameter is the field
fluctuation amplitude, $\dbb$. A lognormal PDF still
appears for small field 
fluctuations (generally the case for {\it large mean fields}), but a
significant low-density excess appears at large fluctuation amplitudes 
({\it weak mean fields}), similar to the behavior at $\gamma > 1$ of
polytropic flows. We interpret these results in terms of simple
nonlinear MHD waves, for which the magnetic pressure behaves linearly
with the density in the case of the slow mode, and quadratically in
the case of the fast wave. Finally, we discuss some implications of
these results, in particular the fact that the effect of the magnetic
field in modifying the PDF is strongest when the mean field is weak.}

\section{Introduction} \label{sec:intro}

A fundamental feature of compressible turbulence is the formation of
density fluctuations, a property  of central
interest in astrophysics, as density excesses (``clouds'') in the
turbulent interstellar medium (ISM) exhibit numerous statistical
properties over a range of
sizes\cite{larson81,blitz93} whose origin is not yet well
understood. Although a 
full understanding of star formation requires knowledge of the full
(multiple-point) statistics in order to determine mean densities as a
function of region size, a first step towards this goal is the
description and physical understanding of the one-point statistics, or
{\it probability density function} (``PDF'') of the mass density field.

Previous studies have shown that the density PDF in turbulent
compressible flows has a lognormal form in the isothermal
case,\cite{VS94,PNJ97,OGS99,OSG00} but exhibits a power-law tail at 
densities larger (smaller) than the mean for flows with
polytropic exponents smaller (larger) than
unity.\cite{scalo_etal98,PVS98,NP99} It should be noted that some of
those works referred to 
purely hydrodynamic flows,\cite{VS94,PVS98} while the rest
referred to magneto-hydrodynamic (MHD) flows.

In particular, ref.\ [\cite{PVS98}] presented a heuristic model for the
development of the lognormal and power-law PDFs, based essentially on
the behavior of the speed of sound with density in those types of
flows. In this sense, the density PDF is a diagnostic for the
dependence of the pressure with density. However, such model applied
only to non-magnetic flows, while the real ISM is most likely
magnetized to a significant extent. In this paper we present
preliminary results on the density PDF of turbulent magnetized flows
in ``1+2/3'' dimensions, as a first attempt to characterize the
behavior of magnetic pressure with density in the fully turbulent
regime. Previous works have focused on the pressure produced by
weakly nonlinear
Alfv\'en waves\cite{MZ95,Spangler}, but here we consider fully
turbulent regimes 
with arbitrarily large magnetic fluctuation amplitudes.

%The outline of the paper is as follows: in \S \ref{sec:HDmod} we briefly
%review the results for the non-magnetic case (\cite{PVS98}), in
%particular the proposed origin of the lognormal and power-law forms of
%the PDF in the isothermal and general polytropic cases. 
%We then present numerical results for the magnetic case,
%in a slab geometry in \S \ref{sec:num_res}, and a proposed
%understanding in terms of the propagation of fast and slow
%finite-amplitude waves in \S
%\ref{sec:waves}. Finally, in \S \ref{sec:concl} we present a summary
%and a discussion of the astropysical implications of our results.

\section{The Non-Magnetic Case} \label{sec:HDmod}

The form of the density PDF in non-magnetic flows and its relation to
the effective equation of state of the system has been understood in
terms of a heuristic model by Passot and \VS,\cite{PVS98} herafter
PVS98 (see also ref.\ [\cite{NP99}]). 
As in ref.\ [\cite{VS94}], this model idealizes the generation
of turbulent density fluctuations as a ``multi-step'' process, in which
the local density at any given point in the flow is the result of a
series of jumps due to the continuous passage of shock waves. 
The generation of densities is an iterative multiplicative
process, in which every new density is obtained through a jump from
the previous one, giving for the final density $\rho_{\rm f}$ in
terms of the intermediate steps $\rho_{i}$:
$\rho_{\rm f} = \rho_0 \prod_i \bigl(\rho_{i+1}/\rho_i\bigr)$.
In the isothermal case, the density jump
is given by $M^2$, where $M$ is the Mach number of the shock and,
because the speed of sound is constant, a given {\it velocity} jump is
always characterized by the same Mach number, regardless of the local
density, so that individual jumps can be regarded as independent, but
extracted from the same distribution. In terms of the variable $s=\ln \rho$,
the iterative process is {\it additive} so that the Central 
Limit Theorem can be applied in the limit of a large number of jumps,
leading to a normal distribution for $s$ and thus a lognormal 
distribution for $\rho$, as observed in the numerical experiments
mentioned above. 

Concerning the variance of the distribution, PVS98 have suggested
(and confirmed numerically), through an analysis of the shock and
expansion waves in the system, that for a large range of Mach numbers
the typical size of the logarithmic jump is expected to be $\sigma_s
\sim \Mrms$, where $\Mrms$ is the rms Mach number.
The mean $s_0$ of the distribution can be directly evaluated from the
mass conservation condition $\langle \rho \rangle
=\int_{-\infty}^{+\infty} e^s P(s) ds = 1$, where $P(s)$ is the PDF of
$s$, yielding $s_0=-\sigma_s^2/2$.
The isothermal model PDF for $s$ thus reads
\begin{equation}
P(s)ds=\frac{1}{\sqrt{2\pi \sigma_s^2}}\exp\bigl[-\frac{(s-s_o)^2}
{2\sigma_s^2}\bigr] ds, \label{eq:isoth_PDF}
\end{equation}
with $\sigma_s^2=\beta \Mrms^2$
and $\beta$ a proportionality constant of order unity. 
%Fig.\ \ref{fig:isoth_PDF}
%illustrates these results, showing the density PDFs for three
%one-dimensional (1D) hydrodynamical simulations with various rms
%Mach numbers. The label $M$ of the curves in fig.\ \ref{fig:isoth_PDF}
%is the Mach number of the velocity unit in the simulations, and is a
%parameter that appears in the nondimensionalization of the equations,
%dividing the pressure-gradient term. This parameter does not coincide
%with the rms Mach number $\Mrms$ of the simulation, but is
%proportional to it, provided the same forcing is applied in all
%cases. The figure
%clearly shows that the maximum of the PDF is displaced towards smaller
%values of $s$ and that the width of the curve increases as the Mach
%number increases. PVS98 showed numerically that eqs.\ (\ref{eq:s_o})and
%(\ref{eq:sigma}) hold in their 1D simulations.
%
%\begin{figure}[t]
%\epsfxsize=20pc % 
%\epsfbox{fig1.ps} % postscript image file name
%\caption{Density PDFs of three isothermal non-magnetic simulations
%with different values of the parameter $M$. The peak of the curve
%moves towards smaller densities and the width of the curve increases
%as $M$ is increased, but the lognormal form is preserved. The dashed
%lines indicate lognormal least-squares fits to the curves.}
%\label{fig:isoth_PDF}
%\end{figure}

In the general polytropic case where the
pressure $P$ behaves as $P \propto \rho^\gamma$, with $\gamma$  the
polytropic exponent, the local Mach number $M(s)$ at a density
$\rho=e^s$ is related to the one at the mean density ($M$) by
$M(s)=M\exp^{(1-\gamma)s/2}$, suggesting an ansatz where the PDF keeps
the same dependence on $M$, provided the above replacement is
made. After relocating the term in $s_0$ from inside the exponential
function to the normalization constant, the model PDF for
the polytropic case reads
\begin{equation}
P(s;\gamma)ds=C(\gamma) \exp \Bigl[\frac{-s^2 e^{(\gamma-1)s}}{2M^2} -
\alpha(\gamma)s\Bigr] ds. \label{eq:PDFgne1}
\end{equation}
\begin{figure}[t]
\epsfxsize=20pc 
\epsfbox{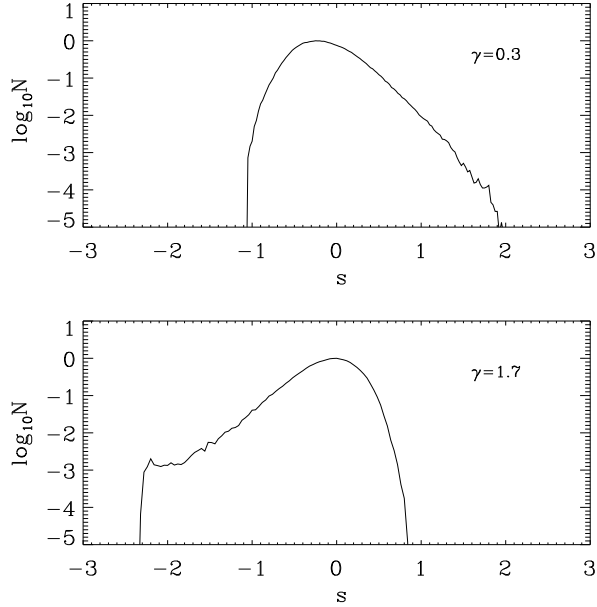}
\caption{Density PDFs for two 1D simulations of polytropic
turbulence, one with $\gamma=0.3$ (left), and the other with
$\gamma=1.7$ (right), both with $M=3$.}
\label{fig:PDFs_gne1}
\end{figure}
This equation shows that when $(\gamma-1)s<0$, the PDF asymptotically
approaches a power law, while in the opposite case it decays faster
than a lognormal. Thus, for $0 <\gamma <1$, the PDF approaches a power
law at large densities ($s>0$), and at low densities ($s<0$) for $\gamma
>1$.

\section{The Magnetic Case}

In what follows we
restrict ourselves to the isothermal case ($\gamma=1$) and to a
propagation
along a uniform ambient magnetic field $\Bo$ and
concentrate on the deviations from the corresponding lognormal PDF
induced by the magnetic field.

\subsection{Numerical Results}  \label{sec:num_res}

The numerical simulations solve the MHD equations using a
pseudo-spectral method in a slab geometry (``1+2/3D'')
(variability is considered only with respect to the spatial
variable $x$ for the three components of the velocity and magnetic
fields).
A resolution of 2048 grid points is used allowing
to handle large enough Mach and Reynolds
numbers with only regular second-order viscosity.
A random acceleration is applied on the $y$- and $z$-components of the
velocity field, i.e.\ transverally to $\Bo$, thus inducing Alfv\'en
waves into the flow. Since these waves have finite amplitudes, they in
turn induce fast and slow magnetosonic waves. The
simulations are evolved over long times (several tens to a few hundred
crossing times at the rms flow velocity) in order to obtain meaningful
statistics for the density PDFs. 

The simulations are essentially characterized by the sonic and
Alfv\'enic Mach numbers, respectively defined 
as $\Ms\equiv u/\cs$ and
$\Ma \equiv u/\va$, where $u$ is the rms velocity,
$\cs$ the sound speed and $\va \equiv \Bo^2/\rho$ the Alfv\'en speed.
In order to investigate the effect of the magnetic field exclusively, we
consider two runs (denoted I and II) with approximatively 
the same value of $\Ms$ (4.06 and 3.74 respectively) but very
different values of  $\Ma$ (0.36 and 1.66 resp.).
These quantities are integrated
over the duration of the run, excluding the first few time units in
order to avoid including the uniform-density initial conditions.
The rms relative field fluctuations $\dbb$ are for runs I and II
respectively 0.75 and 6.65.

%\begin{table}[t]
%\caption{Run parameters\label{table1}}
%\begin{center}
%\footnotesize
%\begin{tabular}{|c|c|c|c|}
%\hline
%Run     & $\Ms$  & $\Ma$ & $\dbb$\\
%\hline
%
%I    &  4.06   &  0.36  &  0.75 \\
%II    &  3.74   &  1.66  &  6.65 \\
%\hline
%
%\end{tabular}
%\end{center}
%\end{table}

\begin{figure}[t] 
\epsfxsize=15pc 
\epsfbox{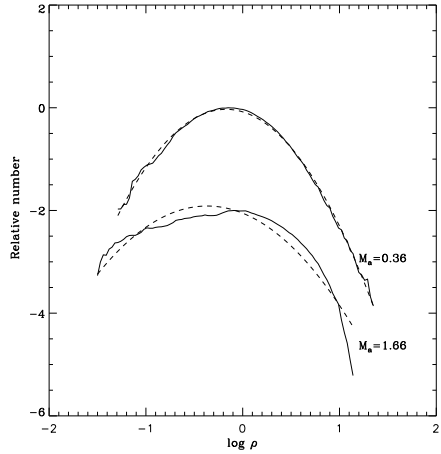}
\caption{Density PDFs of two magnetic simulations with similar values
of the sonic Mach number $\Ms$ but different values of the Alfv\'enic
Mach number $\Ma$. The simulation with small $\Ma$ (large mean field)
is seen to have a nearly lognormal PDF, while the one with large $\Ma$ 
(weak mean field) is seen to have a large excess at low densities,
indicative of an effective ``magnetic'' polytropic exponent larger
than unity.}
\label{fig:mag_PDFs}
\end{figure}

In fig.\ \ref{fig:mag_PDFs} we show the PDFs for the two magnetic
runs, together with lognormal fits (dashed lines). It is clearly seen
that, contrary to what one might expect, the high-$\Ma$ run, which
should be closest to the pure-hydro case, exhibits a 
strong deviation from lognormality, while the low-$\Ma$ PDF is quite
close to a lognormal. This implies that the case with the weakest mean 
field is the one with the strongest effect of magnetic pressure on the 
density PDF.

\subsection{Interpretation in Terms of Simple MHD waves}
\label{sec:waves}

A preliminary interpretation of the behavior reported in the previous
section can be given in terms of the so-called ``simple MHD waves''
(see, e.g., ref.\ [\cite{mann95}]). These are the finite-amplitude
equivalent of linear MHD waves, and have the same three well-known
modes: Alv\'en, slow and fast. A simple wave refers to a solution
that depends only on a single variable, any combinations of the
spatial ($x$) and temporal ($t$) independent variables. However, as
mentioned
above, simple waves have finite amplitudes, and in general 
they develop shocks in a finite time.

A number of properties of simple waves has been reported by
Mann\cite{mann95} (see also Jeffrey and Taniuti\cite{Jeffrey}). 
Among them, most relevant for the present
discussion are the propagation velocities of the three modes, and the
relations between the density and magnetic field fluctuations.
The Alfv\'en mode with speed $V$ such that $V^2 = {\va}_x^2$ is not
associated
with density fluctuations.
The fast and slow speed are given by
\begin{equation}
V_\pm^2 = \frac{{\va}^2+{\cs}^2}{2} \Bigl\{ 1 \pm \bigl[1 -
\frac{4{\cs}^2 {\va}_x^2} {({\va}^2+{\cs}^2)^2}\bigr]^{1/2} \Bigr\}
\label{eq:vpm}
\end{equation}
and the relation between the magnitude of the magnetic field $b$
and the density is given by
\begin{equation}
\frac{db}{d\rho} = \frac{V^2-{\cs}^2}{b},
\label{eq:bvsrho}
\end{equation}
where ${\va}_x^2=b_x^2/\rho$, $\va^2 = b^2/\rho$, and $\cs$
is the sound speed. An important point to note is that eqs.\
(\ref{eq:vpm}) and (\ref{eq:bvsrho}) imply a positive correlation
between $b$ and $\rho$ for the fast mode ($V_+$) and an
anticorrelation for the slow mode ($V_-$),\cite{mann95}.

Inserting eq.\ (\ref{eq:vpm}) in eq.\ (\ref{eq:bvsrho}), 
one can find the density dependence of the magnetic
pressure. Excluding the case where the
$\beta$ of the plasma is of order unity with at the same time
small to moderate field distorsions, this dependence simplifies and
reads (general numerical solutions have been presented by
Mann\cite{mann95}):
\bea
\frac{b^2}{b_x^2} &\approx& a_1 - a_2 \frac{\cs^2 \rho}{b_x^2}
~~~~~~~({\rm slow\ mode}) \label{eq:slow} \\
b^2 &\approx& 2 a_3 \rho^2 ~~~~({\rm fast\ mode}),
\label{eq:fast}
\eea
where $b^2 = b_x^2 + b_\perp^2$, with $b_\perp^2 = b_y^2 + b_z^2$ being
the magnitude of the magnetic field fluctuation. Note that in our
1+2/3 geometry, $b_x (=\Bo)$ is constant. The quantities $a_i$ denote
integration constants. 

From these relations, it can be seen that the magnetic pressure,
$\propto b^2$, scales roughly linearly with the density (albeit
inversely) in the case of the slow mode, but quadratically in the case 
of the fast mode. Note also that, in the case of small field
fluctuations (most often the case if $\beta \equiv \cs^2/(b_x^2/\rho)
\ll 1$) the coefficient of $\rho$ in eq.\ (\ref{eq:slow}) is
small, implying that, for the slow mode, large density fluctuations
can occur even for small variations of $b$. Thus, {\it the slow mode
is expected to dominate density fluctuation production in the case of
small field fluctuations}. Instead, at large enough field
fluctuations, the quadratic dependence of $b^2$ on $\rho$ of the fast
mode eventually overwhelms the linear dependence of the slow mode (and
also of the thermal pressure), so {\it the fast mode is expected
to dominate the density fluctuation production}.

The dependence of the magnetic pressure on the density seems to
describe the observations of sec.\ \ref{sec:num_res} adequately because,
according to the results of the non-magnetic case, a pressure that
depends linearly on the density produces a lognormal density PDF,
while one with a higher-than-linear dependence produces a near power
law at low densities. 

\section{Conclusions} \label{sec:concl}

\subsection{Summary} \label{sec:summary}

In this paper we have reviewed our previous results\cite{PVS98} on the
development of the density PDF as a consequence of the effective
(polytropic) equation of state, and applied them to an understanding
of the PDF in the magnetic case.

In the context of the model presented in ref. [\cite{PVS98}], 
isothermal non-magnetic flows have lognormal mass density PDFs while in the 
non-isothermal cases, a power-law tail develops at high Mach number, at
either
$\rho > \langle \rho \rangle$ or $\rho < \langle \rho \rangle$
depending on whether $\gamma <1$ or $\gamma > 1$, respectively.

In the magnetic case, we considered only isothermal cases and
a propagation parallel to the ambient magnetic field, but
reported that a deviation from the lognormal PDF occurs {\it when 
the magnetic fluctuations are large}, a case expected when the mean
field is small. We interpreted this effect in terms of so-called
``simple'', finite-amplitude nonlinear MHD waves. Indeed, for the slow 
mode of these waves, the magnetic pressure $b^2$ behaves linearly with
the density $\rho$, while for the fast mode, the magnetic pressure
behaves
quadratically with $\rho$. Together with the fact that for the slow
wave the density is weighted by a small factor, this suggests that the 
production of density fluctuations is dominated by the slow waves at
small field deviations (i.e., $\Bo$ large) giving a lognormal PDF
again, while for weak mean fields (large field deviations) the
quadratic density dependence of $b^2$ produces an excess at small
densities in the PDF, corresponding to the $\gamma > 1$ case of the
polytopic description. A more detailed study, including in
particular the case of non-parallel propagation is in progress.

\subsection{Astrophysical Implications} \label{sec:astro_impl}

The results of this paper have a number of implications in the
astrophysical context. First, we have seen that the main parameter
determining the effect of the magnetic field appears to be the
magnetic fluctuation amplitude, $\dbb$, rather than $\beta$ alone,
which is the parameter most frequently used to characterize MHD
flows. In other words, it is important to have a knowledge of the
relative importance of the turbulent fluctuations (as given by the
Alfv\'enic Mach number) in addition to simply 
the ratio of thermal to magnetic pressures (as given by $\beta$).

Second, as far as the density PDF is concerned, the effect of the
magnetic field is most notorious when the field fluctuations are large 
(generally, when the mean field is weak). This suggests that the limit
of vanishing field does not approach the non-magnetic case, and in
this sense the latter is singular. This may imply that it is
inadequate to model the ISM as a non-magnetic flow even if the degree
of magnetization is very low, as has started to be recently contended
at the level of molecular clouds (see, e.g., ref.\ [\cite{PN99}]).

\section*{Acknowledgements}
Part of this work was completed while the authors enjoyed the
hospitality of the Institute for Theoretical Physics at the University
of California at Santa Barbara, as participants of the Astrophysical
Turbulence Program. This work has received partial funding 
from grants Conacyt (M\'exico) 27752-E to E.V.-S, the CNRS National
Program ``PCMI'' to T.P. and NSF PHY94-07194.

\end{document}